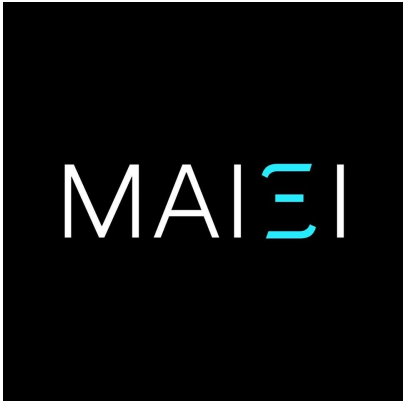

Montreal AI Ethics Institute

*An international, non-profit research institute helping humanity define its place in a world increasingly driven and characterized by algorithms*

Website: https://montrealethics.ai
Newsletter: https://aiethics.substack.com

# Report prepared by the Montreal AI Ethics Institute for the European Commission's Whitepaper on AI 2020
## On Artificial Intelligence - A European approach to excellence and trust

*Based on insights and analysis by the Montreal AI Ethics Institute (MAIEI) staff on the policy documents from the European Commission and supplemented by workshop contributions from the AI Ethics community convened by MAIEI on May 27, 2020 and June 3, 2020*

Primary contacts for the report:

Abhishek Gupta (abhishek@montrealethics.ai)
Founder, Montreal AI Ethics Institute
Machine Learning Engineer, Microsoft

Camylle Lanteigne (camylle@montrealethics.ai)
AI Ethics Researcher, Montreal AI Ethics Institute

## Overview of our recommendations

1. Focus efforts on the research and innovation community, member states, and the private sector, as well as those that should come first in Europe's AI strategy.
2. Create alignment between the major trading partners' policies and the EU policies governing the development and use of AI.
3. Analyze the gaps in the current ecosystem between theoretical frameworks and approaches to building trustworthy AI systems to create more actionable guidance that helps organizations implement these principles in practice.
4. Focus on coordination and policy alignment, particularly in two areas: increasing the financing for AI start-ups and developing skills and adapting current training programs.
5. Focus on mechanisms that promote private and secure sharing of data in the building up of the European data space, leveraging technical advances like federated learning, differential privacy, federated analytics, and homomorphic encryption.
6. Create a network of existing AI research excellence centres to strengthen the research and innovation community, with a focus on producing quality scholarship work that takes into account a diverse array of values/ethics.
7. Promote knowledge transfer and develop AI expertise for SMEs as well as support partnerships between SMEs and the other stakeholders through Digital Innovation Hubs.
8. Add nuance to the discussion regarding the opacity of AI systems, so that there is a graduated approach to how these systems are governed and in which place there is a requirement for what degree of explainability and transparency.
9. Create a process for individuals to appeal an AI system's decision or output, such as a 'right to negotiate,' which is similar to the 'right to object' detailed in the General Data Protection Regulation (GDPR).
10. Implement new rules and strengthen existing regulations to better address the concerns regarding AI systems.
11. Ban the use of facial recognition technology, which could significantly lower risks regarding discriminatory outcomes and breaches in fundamental rights.
12. Hold all AI systems (e.g. low-, medium-, and high-risk applications) to similar standards and compulsory requirements.
13. Ensure that if biometric identification systems are used, they fulfill the purpose for which they are implemented while also being the best way of going about the task.
14. Implement a voluntary labelling system for systems that are not considered high-risk, which should be further supported by strong economic incentives.
15. Appoint individuals to the human oversight process who understand the AI systems well and are able to communicate any potential risks effectively with a variety of stakeholders so that they can take the appropriate action.

**Section 1: an ecosystem of excellence**

1. **In your opinion, how important are the six actions proposed in section 4 of the White Paper on AI (1-5: 1 is not important at all, 5 is very important)?**

Working with member states: **4**
Focussing the efforts on the research and innovation community: **5**
Skills: **4**
Focus on SMEs: **3**
Partnership with the private sector: **4**
Promoting the adoption of AI by the public sector: **2**

For this question, we consider the first four areas mentioned to be of some importance in building an ecosystem of excellence. However, we believe that focussing the efforts on the research and innovation community, working with member states, and partnering with the private sector to be the most important aspects, as well as those that should come first in Europe's AI strategy. Indeed, by focusing on these three actions, we believe AI implementation throughout the European Union will be most efficient. Focusing the efforts on the research and innovation will allow the EU to build strong AI expertise and leadership within its borders, and develop safe and powerful AI. Getting this component right seems crucial to us. Similarly, working with member states will create a uniform AI strategy across the EU, avoiding a potentially fragmented and difficult to regulate implementation of AI across countries. As a starting point creating alignment between the major trading partners' policies and the EU policies governing the development and use of AI will have immediate economic[1] and intellectual property rights implications, without which the EU risks becoming an island in the research and development of AI systems.

Partnering with the private sector will allow the EU to benefit from additional input and investment, while also helping ensure the private sector and the state see eye to eye in terms of how AI is to be implemented. Additionally, we believe that it is of paramount importance to consider the carbon impacts of AI systems, especially developing standardized measurement mechanisms[2] that help to facilitate comparisons across solutions developed by different member states. This will help both developers and consumers pick, based on informed and uniform metrics, solutions that uphold European values.

We don't consider promoting the adoption of AI by the public sector to be a terribly important action for the EU, at least right now. We agree with Fanny Hidvegi and Daniel Leufer when they write: "The uptake of any technology, particularly in the public sector, should not be a standalone goal and it is not of value in itself. [...] In cases where there are no serious negative impacts and there is evidence of real benefit, AI-based systems can be considered as an option alongside other approaches, but we must ensure that policy makers are not led astray by marketing slogans and unfounded AI hype".[3] A significant concern in advocating for the adoption of AI in a broad-based manner by the public sector within the healthcare and transport industry as it stands at the moment are the vast unaddressed failure modes in AI systems. As an example, machine learning security[4] is an emergent field where new attack surfaces exposed as a function of integrating machine learning into existing software infrastructure opens up new vulnerabilities that are left underprotected by traditional cybersecurity measures. The awareness within the machine learning community on these potential failures is still in the nascent stages and requires deeper analysis and integration into the AI development and deployment lifecycle before the risks are mitigated to a degree where these systems can be considered safe from a widespread deployment perspective.

---

[1] Gupta, A. (2020). Montreal AI Ethics Institute's Response to Scotland's AI Strategy. arXiv preprint arXiv:2006.06300
[2] Gupta, A., Lanteigne, C., & Kingsley, S. (2020). SECure: A Social and Environmental Certificate for AI Systems. arXiv preprint arXiv:2006.06217.

From a public procurement perspective, it is important that the procurement officers and regulators are equipped and empowered with the necessary tooling and techniques[5] to evaluate the trustworthiness of the AI systems which might be used in diverse ways in downstream tasks. As an example, asking vendors to provide verifiable claims[6], and other supporting documentation taking the form of datasheets for datasets[7] that were used to train the AI systems, model cards for model reporting[8], and nutrition labels for datasets[9].

One of the phrases used in the whitepaper, that might have unintentionally been used, relates to the "... race for global leadership is ongoing …" should be framed and expressed differently. From a semantic perspective, it sets up the global AI development ecosystem in a competitive manner whereby there might be strong incentives to restrict the free-flow of talent and knowledge across the world and within Europe which could hinder the development of AI, especially as it relates to responsible AI development and deployment. Specifically, such a competitive framing has the potential to create incentives where there might be AI development and deployment that is done to optimize for metrics[10], Goodhart's law[11], rather than achieving the high-quality of AI systems that respect the fundamental rights and values of European citizens. Additionally, it also has the potential to negatively impact the "brand" of European solutions which are today held up as a benchmark for building systems that consider deeply the societal and environmental impacts.

---

[3] Hidvegi, F., & Leufer, D. (2020, June 11). Trust and excellence—The EU is missing the mark again on AI and human rights. *Access Now*.
https://www.accessnow.org/trust-and-excellence-the-eu-is-missing-the-mark-again-on-ai-and-human-rights/

[4] Leong, B. (2019, September 20). *Warning Signs: Identifying Privacy and Security Risks to Machine Learning Systems*. Future of Privacy Forum.
https://fpf.org/2019/09/20/warning-signs-identifying-privacy-and-security-risks-to-machine-learning-systems/

[5] Varner, M., & Sankin, A. (2020, February 25). *Suckers List: How Allstate's Secret Auto Insurance Algorithm Squeezes Big Spenders – The Markup*. The Markup. Retrieved 14 June 2020, from https://themarkup.org/allstates-algorithm/2020/02/25/car-insurance-suckers-list

[6] Brundage, M., Avin, S., Wang, J., Belfield, H., Krueger, G., Hadfield, G., ... & Maharaj, T. (2020). Toward trustworthy AI development: mechanisms for supporting verifiable claims. arXiv preprint arXiv:2004.07213.

[7] Gebru, T., Morgenstern, J., Vecchione, B., Vaughan, J. W., Wallach, H., Daumeé III, H., & Crawford, K. (2018). Datasheets for datasets. arXiv preprint arXiv:1803.09010.

[8] Mitchell, M., Wu, S., Zaldivar, A., Barnes, P., Vasserman, L., Hutchinson, B., ... & Gebru, T. (2019, January). Model cards for model reporting. In Proceedings of the conference on fairness, accountability, and transparency (pp. 220-229).

[9] Holland, S., Hosny, A., Newman, S., Joseph, J., & Chmielinski, K. (2018). The dataset nutrition label: A framework to drive higher data quality standards. arXiv preprint arXiv:1805.03677.

From a skills and training perspective, while we appreciate the explicit mention of increasing awareness about AI systems, we advocate that it is equally important to consider some of the complementary areas in the social sciences as a focus, an emphasis on learning to learn, continuous learning, and other measures that better prepare the workforce for the future of work. Rapid automation penetrating different industries at skill levels means that the future of work is inherently more stochastic as the pace of innovation and technology deployment quickens. From an educational and skills development perspective we foresee that it is hard to predict what might be the relevant skills in the future, but an emphasis on empowering the people to pick up skills in a continual manner, providing them with necessary societal and institutional support so that they are able to successfully transition into new functional roles as needed.

From an inclusion perspective, a focus on equipping more women trained and employed in AI is of utmost importance, we would also like to see language around the inclusion of other underrepresented groups, lest there is a risk of reduced attention paid to truly empowering all EU citizens, regardless of their backgrounds, lived experiences, and other characteristics, in participating to shape the development and deployment of AI systems[12].

One of our key recommendations is to analyze the gaps in the current ecosystem such that we can bridge the gap between theoretical frameworks and approaches to building trustworthy AI systems to more actionable guidance that helps organizations implement these principles in practice. From our work, we have found that to be a major roadblock for organizations that are trying to adopt these principles at a high-level but finding themselves in a situation where they are not well-prepared to tangibly apply them to the products and services that they are developing. We highlighted some of these challenges and potential solutions from a legal and regulatory perspective in our submission that we made to the Office of the Privacy Commissioner of Canada[13].

---

[10] Thomas, R., & Uminsky, D. (2020). The Problem with Metrics is a Fundamental Problem for AI. arXiv preprint arXiv:2002.08512.

[11] Reynaert, M., & Sallee, J. (2016). *Corrective Policy and Goodhart's Law: The Case of Carbon Emissions from Automobiles* (No. w22911; p. w22911). National Bureau of Economic Research. https://doi.org/10.3386/w22911

[12] Executive Office of the President. (2016). Artificial Intelligence, Automation, and the Economy. https://obamawhitehouse.archives.gov/sites/whitehouse.gov/files/documents/Artificial-Intelligence-Automation-Economy.PDF

*Are there other actions that should be considered? (500 chars max)*

*Please see the attached PDF for more details*

2. **In your opinion, how important is it in each of these areas to align policies and strengthen coordination as described in section 4.A of the White Paper (1-5: 1 is not important at all, 5 is very important)?**

Strengthen excellence in research **3**
Establish world-reference testing facilities for AI **2**
Promote the uptake of AI by business and the public sector **4**
Increase the financing for start-ups innovating in AI **5**
Develop skills for AI and adapt existing training programmes **5**
Build up the European data space **4**

Coordination and policy alignment seem the most crucial to us in two areas: increasing the financing for AI start-ups and developing skills and adapting current training programs. We highlight this as the most important because we believe that disparities in these areas would be detrimental to economic and opportunity equality within the EU. Without strong start-up funding, it seems highly plausible that only a few AI hubs will form, leaving many cities behind in terms of economic development and opportunities to benefit from the AI economy. Disparities in terms of AI skills and training across EU countries may also lead those who are hoping to pursue a career in AI to congregate to a few select areas, draining other towns and cities of their qualified and talented young people, which will negatively affect their economy. The clustering effects take place naturally through the creation of hubs that tend to concentrate talent in a small geographic region to leverage the benefits[14] of working in close proximity and capitalize on supporting infrastructure. Yet, as mentioned above, there is a tremendous potential for this to create problems in terms of unequal distribution of opportunities and unnecessary strain on infrastructure in some regions while leaving others woefully behind. Inclusive growth requires the distribution of opportunities across regions and in the designing of a new ecosystem that seeks to leverage the EU's diversity in terms of backgrounds and experiences, it would be crucial to be "inclusive-by-design" to support the localized addressal of problems and creation of solutions that take the cultural and contextual sensitivities of different regions into consideration.

---

[13] Snyder Caron, M., & Gupta, A. (2020). *Response to Office of the Privacy Commissioner of Canada Consultation Proposals pertaining to amendments to PIPEDA relative to Artificial Intelligence.* Montreal AI Ethics Institute.
https://montrealethics.ai/wp-content/uploads/2020/03/FINAL-DOC-Submission-to-OPC-consultation.pdf
[14] Hsieh, C. T., & Moretti, E. (2015). Why do cities matter? Local growth and aggregate growth.

From the testing facilities perspective, while we believe that this is an excellent initiative, as of yet, there aren't sufficient mechanisms that can do so in a standardized manner that facilitate meaningful comparison and hence we rate the priority of that being lower. There are some existing initiatives that seek to utilize benchmarks to assess the technical robustness of AI systems[15] but they are preliminary works from the perspective that they assess the performance of AI systems on a narrow range of measures and don't yet take a holistic approach to evaluating the trustworthiness of AI systems.

In the building up of the European data space, we would like to call attention to focussing on mechanisms that promote private and secure sharing of data, leveraging on technical advances like federated learning[16], differential privacy[17], federated analytics[18], and homomorphic encryption[19] (though we acknowledge that homomorphic encryption as of yet only works for a limited set of computations, it nonetheless presents an exciting opportunity). In particular, within the centers of excellence, an emphasis placed on practical integration of these techniques within the context of creating a strong data ecosystem within the EU that allows for seamless data sharing which is crucial for building large repositories of high-quality data[20] that can have implications in terms of the quality of supervised machine learning systems that are trained.

*Are there other actions that should be considered? (500 chars max)*

*Please see the attached PDF for more details*

3. **In your opinion how important are the three actions proposed in sections 4.B, 4.C and 4.E of the White Paper on AI (1-5: 1 is not important at all, 5 is very important)?**

Support the establishment of a lighthouse research centre that is world class and able to attract the best minds **3**
Network of existing AI research excellence centres **5**
Set up a public-private partnership for industrial research **4**

---

[15] Martínez-Plumed, F., & Hernández-Orallo, J. (2018). Analysing Results from AI Benchmarks: Key Indicators and How to Obtain Them. arXiv preprint arXiv:1811.08186.
[16] Yang, T., Andrew, G., Eichner, H., Sun, H., Li, W., Kong, N., ... & Beaufays, F. (2018). Applied federated learning: Improving google keyboard query suggestions. arXiv preprint arXiv:1812.02903.
[17] Dwork, C., & Roth, A. (2014). The algorithmic foundations of differential privacy. Foundations and Trends in Theoretical Computer Science, 9(3-4), 211-407.
[18] Ramage, D., & Mazzocchi, S. (2020, May 27). Federated Analytics: Collaborative Data Science without Data Collection. *Google AI Blog*.
http://ai.googleblog.com/2020/05/federated-analytics-collaborative-data.html
[19] Naehrig, M., Lauter, K., & Vaikuntanathan, V. (2011, October). Can homomorphic encryption be practical?. In Proceedings of the 3rd ACM workshop on Cloud computing security workshop (pp. 113-124).
[20] Sessions, V., & Valtorta, M. (2006). The Effects of Data Quality on Machine Learning Algorithms. ICIQ, 6, 485-498.

In this case, creating a network of existing AI research excellence centres seems to us the most important actions of the three presented above. This action also seems more likely than a lighthouse research centre to attract very talented individuals in the field of AI. This is because, for one, a single research centre can only welcome so many researchers, as compared to a whole network. In addition, a network of research centres can cover more ground much more efficiently. A network is also more robust than a single research centre, and is more likely to be able to make efficient use of funding, especially considering the large amounts of money being invested in the field of AI currently. Also, pursuant to our prior answers, we believe a distributed approach is best for fostering more opportunities and creating a more equitable distribution of the gains to be had from AI deployment in the EU. Specifically, we advocate a pan-European approach borrowing from the work happening in Canada with the establishment of Mila, Vector Institute, and the Alberta Machine Intelligence Institute which has helped to build up multiple centers with a diverse set of expertise sourced from different parts of the country. This will be quite important given the diversity of values and cultures within the EU which should be accommodated while finding a common thread that unites these efforts through the shared European values.

Given the rapid pace of moving research in the AI domain from theoretical, lab experiments in various organizations to concrete use in the products and services in the real-world, it is essential that there is synchronization between the public and private organizations so that they keep each other apprised of the latest changes and potential risks from the deployment of solutions can be caught early in the cycle. Additionally, given that a lot of the top talent in the AI ends up working at private corporations[21], there are tremendous gains to be realized if there is an instrument that opens up an exchange of knowledge between public and private industry through both formal and informal channels.

To strengthen the research and innovation community, there should be a focus on producing scholarship work that avoids the pitfalls that emerging fields experience when they is a high degree of attention on it and an influx of people working on it. Some of those concerns are highlighted in the work from Lipton and Steinhardt[22] which gives guidance to both new and experienced researchers on how they can collectively improve the quality of work done in the Ai domain. We advocate that the EC take this into consideration when putting in place various instruments that will shape the AI ecosystem in Europe.

*Are there any other actions to strengthen the research and innovation community that should be given a priority? 500 character(s) maximum*

*Please see the attached PDF for more information*

---

[21] Roca, T. (2019). AI talent in the European labour market.
[22] Lipton, Z. C., & Steinhardt, J. (2018). Troubling trends in machine learning scholarship. arXiv preprint arXiv:1807.03341.

4. **In your opinion, how important are each of these tasks of the specialised Digital Innovation Hubs mentioned in section 4.D of the White Paper in relation to SMEs (1-5: 1 is not important at all, 5 is very important)?**

Help to raise SMEs' awareness about potential benefits of AI **2**
Provide access to testing and reference facilities **3**
Promote knowledge transfer and support the development of AI expertise for SMEs **5**
Support partnerships between SMEs, larger enterprises and academia around AI projects **5**
Provide information about equity financing for AI startups **3**

In our view, Digital Innovation Hubs should focus on promoting knowledge transfer and developing AI expertise for SMEs as well as supporting partnerships between SMEs and the other stakeholders mentioned above. Compared to other areas Digital Innovation Hubs could focus on, these two seem to be where maximum impact can be achieved. This greater impact because first, the two aforementioned areas seem to be the ones SMEs are most likely to struggle with if they do not receive support from Digital Innovation Hubs and the EU. Other tasks seem more easily attainable for SMEs, even with little or no government support. Furthermore, knowledge transfer and AI expertise within SMEs appears particularly challenging considering that large technology and AI companies are often much more attractive to those in the field of AI. Bigger companies can for example offer larger salaries, more perks, and more prestige or career capital. This leaves SMEs in a difficult position when they attempt to build AI expertise for their business. Thus, it is crucial that Digital Innovation Hubs help bridge the gap between SMEs and large companies in terms of AI expertise, as it seems unlikely to get better for SMEs without help from the EU. Additionally, given the dominance of the current supervised machine learning paradigm we also see creating data commons[23] or another data pooling instrument that allows for smaller firms to access high-quality data in large magnitudes which are essential for the training of complex models. Additionally, access to common compute infrastructure will also be essential in enabling SMEs to be competitive in a marketplace where complex models that require large amounts of data are also inaccessible for smaller firms to train because of financial and other constraints.

We also see the Digital Innovation Hubs as the perfect instrument for disseminating standards and tangible guidance on how to operationalize theoretical approaches in responsible AI to something that practitioners can implement into their everyday work. Making responsible AI the norm[24] rather than the exception will be one of the key values that needs to be put forth as a tenet of the Digital Innovation Hubs. Large and small organizations struggle equally when it comes to putting principles into practice. One of the biggest hurdles in adoption of principles and putting them into practice is that there isn't widespread awareness of the technical options that designers and developers have to operationalize the principles. The Hub can act as a central repository for sharing best practices and compile feedback and insights from the real=world deployment of responsible AI framework and help people make better choices in building responsible AI systems choosing more effective frameworks and tools.

---

[23] Miller, P., Styles, R., & Heath, T. (2008). Open Data Commons, a License for Open Data. LDOW, 369.
[24] *About*. (n.d.). Montreal AI Ethics Institute. Retrieved 14 June 2020, from https://montrealethics.ai/about/.

*Are there any other tasks that you consider important for specialised Digital Innovations Hubs? 500 character(s) maximum*

*Please see the attached PDF for more information.*

**Section 2: An ecosystem of trust**

5. **In your opinion, how important are the following concerns about AI (1-5: 1 is not important at all, 5 is very important)?**

AI may endanger safety **5**
AI may breach fundamental rights (such as human dignity, privacy, data protection, freedom of expression, workers' rights etc.) **5**
The use of AI may lead to discriminatory outcomes **5**
AI may take actions for which the rationale cannot be explained **5**
AI may make it more difficult for persons having suffered harm to obtain compensation **5**
AI is not always accurate **4**

When talking about the opacity of AI systems, there is a need to add nuance to the discussion so that there is a graduated approach to how these systems are governed and in which place there is a requirement for what degree of explainability[25] and transparency. Specifically, in sectors that are highly regulated, for example finance and healthcare, there are a class of models like binary decision trees that offer a greater degree of transparency in terms of how decisions were made, compared to the more sophisticated techniques like neural networks which can suffer from the black-box effect. Opacity as a label applied to AI oversees these nuances and creates issues in terms of potentially creating regulatory requirements that are misdirected.

While we consider accuracy in AI systems to be crucial, we are somewhat less concerned with this feature than the others mentioned for the following reasons. First, the other concerns mentioned above are more targeted, and highlight more tangible risks. Second, the accuracy of AI systems itself is not the central problem; what follows from an AI system's lack of accuracy, and the context around the failure, are the crucial aspects. For one, inaccurate AI becomes especially significant in the context of "mathwashing",[26] where individuals place unwarranted trust in a system just because it is a numerical, math-based system. Because of mathwashing, individuals don't expect or don't believe that AI systems can be fallible. This can be harmful if humans then continue to use or don't know how to deal with an inaccurate AI system, and can give rise to some of the concerns originally mentioned in the question posed above.

---

[25] Zhou, Y., & Danks, D. (2020, February). Different Intelligibility for Different Folks. In Proceedings of the AAAI/ACM Conference on AI, Ethics, and Society (pp. 194-199).
[26] Woods, T. (2016, June 8). *'Mathwashing,' Facebook and the zeitgeist of data worship*. Technical.ly Brooklyn. https://technical.ly/brooklyn/2016/06/08/fred-benenson-mathwashing-facebook-data-worship/

In addition, "automation surprise"[27] may follow from placing too much trust in AI systems. This is true even when there is a so-called human-in-the-loop, as this human may be a "token human", placed there to watch over the AI system but failing to do so because they do not fully understand how it functions or overestimate the AI system's accuracy.[28] Additionally, AI systems have the potential to gradually build trust from the user in the sense that they are reliable systems for doing predictions when they work. Specifically, automated systems can help smooth out smaller errors to give users a more seamless experience but this opens up the door for more drastic failures that can arise when the human is caught off-guard as the system performs anomalously and leads to more catastrophic failures. There is a strong need for a system that allows for graceful failures[29] in the cases where the AI system is unable to handle potentially out-of-distribution data[30] that can lead to erratic predictions.

When AI systems fail, and when individuals have reasons to question the accuracy of an AI system, it is crucial to have a process for individuals to appeal an AI system's decision or output. We suggest implementing a "right to negotiate". This is not unlike the 'right to object' detailed in the General Data Protection Regulation (GDPR), but the right to negotiate goes one step further: it would guarantee that "the individual concerned will be provided with the reasons and general factors underlying the business decision, automated or not, and can attempt to justify how certain factors should be changed, if he considers these as erroneous or unreasonable, or he may be provided the opportunity to provide new or other data to assist in the business decision to go in his favour".[31]

Thus, the issues surrounding AI's lack of accuracy, and solutions regarding inaccuracies appear to be much more related to the second-order effects of inaccurate AI systems. For systems that rely on reinforcement learning techniques, there is an urgency in evaluating whether such systems are vulnerable to failures, for example the ones specified in the work from Amodei et al.

---

[27] Sarter, N., Woods, D., & Billings, C. (1997). Automation surprises. In Handbook of Human Factors and Ergonomics (Vol. 2, pp. 1926–1943).

[28] Kahn, J. (2020, May 5). *A.I. engineers should spend time training not just algorithms, but also humans*. Fortune. https://fortune.com/2020/05/05/a-i-engineers-should-spend-time-training-not-just-algorithms-but-also-the-humans-who-use-them/

[29] Weber, J., & Wotawa, F. (2010). Combining Runtime Diagnosis and AI-planning in a Mobile Autonomous Robot to Achieve a Graceful Degradation after Software Failures. In ICAART (1) (pp. 127-134).

[30] Ren, J., Liu, P. J., Fertig, E., Snoek, J., Poplin, R., Depristo, M., ... & Lakshminarayanan, B. (2019). Likelihood ratios for out-of-distribution detection. In Advances in Neural Information Processing Systems (pp. 14680-14691).

[31] Snyder Caron, M., & Gupta, A. (2020). *Response to Office of the Privacy Commissioner of Canada Consultation Proposals pertaining to amendments to PIPEDA relative to Artificial Intelligence.* Montreal AI Ethics Institute. https://montrealethics.ai/wp-content/uploads/2020/03/FINAL-DOC-Submission-to-OPC-consultation.pdf

[32] An additional area of exploration is the domain of machine learning security, specifically as it relates to the robustness of AI systems. Techniques like model inversion, data poisoning, model stealing, and other attacks as defined in machine learning security literature[33] are important to address from the cybersecurity perspective. Traditional cybersecurity measures are insufficient in these scenarios as demonstrated in this vulnerability that was found in ProofPoint's phishing emails classifier[34]. There are many other attacks that might go undetected or unreported and this has serious implications in terms of the reliability and safety of the system in real-world use.

*Do you have any other concerns about AI that are not mentioned above? Please specify: 500 character(s) maximum*

*Please see the attached PDF for information*

6. **Do you think that the concerns expressed above can be addressed by applicable EU legislation? If not, do you think that there should be specific new rules for AI systems?**

Current legislation is fully sufficient
**Current legislation may have some gaps [x]**
There is a need for a new legislation
Other
No opinion

We believe new rules should be implemented to better regulate the concerns expressed in the previous question. As mentioned above, we propose a right to negotiation to help mitigate the fact that AI may make it harder for some individuals to get compensation, which can be due to AI taking actions that cannot be fully explained. Another measure the EU could take is to ban facial recognition technology, which could significantly lower risks regarding discriminatory outcomes and breaches in fundamental rights. Very recently, IBM committed to stopping the development of facial recognition technology due to the risk of racial profiling and the use of facial recognition by law enforcement more broadly.[35] This highlights how it is not too early, contrary to what some believe, to ban facial recognition. We believe that banning facial recognition could be effective in limiting instances where AI violates fundamental rights or acts in a way that is discriminatory.

---

[32] Amodei, D., Olah, C., Steinhardt, J., Christiano, P., Schulman, J., & Mané, D. (2016). Concrete problems in AI safety. arXiv preprint arXiv:1606.06565.

[33] Yuan, X., He, P., Zhu, Q., & Li, X. (2019). Adversarial examples: Attacks and defenses for deep learning. IEEE transactions on neural networks and learning systems, 30(9), 2805-2824.

[34] *CVE Home Page*. (n.d.). Common Vulnerabilities and Exposure. Retrieved 14 June 2020, from https://cve.mitre.org/cgi-bin/cvename.cgi?name=CVE-2019-20634

[35] Hern, A. (2020, June 9). IBM quits facial-recognition market over police racial-profiling concerns. *The Guardian*. https://www.theguardian.com/technology/2020/jun/09/ibm-quits-facial-recognition-market-over-law-enforcement-concerns

What is more, current legislation (not exclusively in the EU, but around the world) is ill-prepared to deal with security threats that stem from AI. Automated weapons, automated cyberattacks, and even AI-generated propaganda are very new and quite different from their predecessors.[36] Hence legislation is not particularly well-adapted to protect citizens against these security threats, nor to prosecute those who are breaking the law using AI.

7. **If you think that new rules are necessary for AI systems, do you agree that the introduction of new compulsory requirements should be limited to high-risk applications (where the possible harm caused by the AI system is particularly high)?**

Yes
**No [x]**
Other
No opinion

We do not believe that new compulsory requirements for AI should be limited to "high-risk" AI systems or applications. While we do support compulsory requirements for AI that is deemed "high-risk", we also believe that AI that may be "medium-risk" or even "low-risk" should also be subject to similar compulsory requirements. This is because AI systems that can initially seem to harbor only medium or low risk can nevertheless place an undue burden on individuals through, for example, discrimination, lack of explainability, or by making it more difficult to appeal a decision. An example of a low-risk application of AI may be identifying the content of photographs in a person's phone (where a picture of a dog would be assigned the label "dog"). This doesn't seem particularly risky at first glance. However, this can have non-negligible consequences, as it did when Google's photograph-labelling algorithm labelled a photograph of a Black man and woman as being of "gorillas".[37] This error, while it does not lead to the wrongful conviction of a person, or one's job application being wrongfully rejected, can still have serious societal impacts by playing into racist stereotypes and unnecessarily affect Black individuals' mental health. We consider these harms to be significant enough to warrant compulsory requirements in the same way explicitly high-risk applications of AI do. Additionally, it also relates to precedent setting whereby having lax regulations around low- and medium-risk applications that can have non-negligible effects when scaled across many users, can create an environment where some of these seemingly innocuous effects can lead to unknown outcomes when the simpler modules are combined to build a more complex systems.

---

[36] Brundage, M., Avin, S., Clark, J., Toner, H., Eckersley, P., Garfinkel, B., Dafoe, A., Scharre, P., Zeitzoff, T., Filar, B., Anderson, H., Roff, H., Allen, G. C., Steinhardt, J., Flynn, C., hÉigeartaigh, S. Ó., Beard, S., Belfield, H., Farquhar, S., … Amodei, D. (2018). The Malicious Use of Artificial Intelligence: Forecasting, Prevention, and Mitigation. *ArXiv:1802.07228 [Cs]*. http://arxiv.org/abs/1802.07228

[37] Lipton, Z. C. (2016, November 7). The Foundations of Algorithmic Bias. *Approximately Correct*. http://approximatelycorrect.com/2016/11/07/the-foundations-of-algorithmic-bias/

To mention only another risk of an image-labeling algorithm like Google's: because the algorithm needs to access individuals' images, there is a risk that these can be used to train other, more risky technology (like facial recognition) without users' consent. There is mounting evidence already of the use of *masked face selfies[38],* where by the system is being trained to use fewer and different features for recognizing faces even a significant part of the face is obscured.

Therefore, we would not recommend that seemingly low or medium risk applications of AI be held to lower standards than explicitly high-risk applications since AI that may, on the surface, seem inoffensive can easily become "high-risk". For these reasons, we do not agree that new compulsory requirements should apply only to "high-risk" applications of AI.

A comment on assessing the risk levels of an automated system, when we have online learning systems that can adapt their behaviour to meet the external metrics that have been served to collect the model's performance. This can surface some potentially innovative behavior from automated systems. Gaining a deeper understanding of how the AI systems operate along with serendipitous collisions across functional roles will help to find ways through which we can prevent specification gaming[39]

*If you wish, please indicate the AI application or use that is most concerning ("high-risk") from your perspective: 500 character(s) maximum*

*Please see the attached PDF for more information.*

8. **In your opinion, how important are the following mandatory requirements of a possible future regulatory framework for AI (as section 5.D of the White Paper) (1-5: 1 is not important at all, 5 is very important)?**

The quality of training data sets **4**
The keeping of records and data **5**
Information on the purpose and the nature of AI systems **2**
Robustness and accuracy of AI systems **5**
Human oversight **3**
Clear liability and safety rules **5**

From a regulatory framework standpoint, the keeping of records and data will be the most essential component since it can help to facilitate and achieve some of the other points

---

[38] Ng, A. (2020, May 19). *Your face mask selfies could be training the next facial recognition tool*. CNET. https://www.cnet.com/news/your-face-mask-selfies-could-be-training-the-next-facial-recognition-tool/

[39] Krakovna, V., Uesato, J., Mikulik, V., Rahtz, M., Everitt, T., Kumar, R., Kenton, Z., Leike, J., & Legg, S. (2020, April 21). *Specification gaming: The flip side of AI ingenuity*. Deepmind. /blog/article/Specification-gaming-the-flip-side-of-AI-ingenuity

mentioned here. It is fundamental in terms of creating traces through the system which can be used to audit the system, providing a concrete mechanism for human oversight where post-hoc, an analysis akin to that made by forensic analysts who use system logs in the field of information security to detect if there were violations of the confidentiality, integrity, and availability of the system. In a previous answer, we mentioned

Safety and liability rules, when expressed very clearly, have the potential to steer progress in the field in a manner that gravitates towards building private- and secure-by-design systems. The goal through doing so would be to provide actionable guidance to the designers and developers so that they can put the principles into practice.

As highlighted in a previous answer, we believe human oversight needs to be done but in a meaningful way so that they don't end up becoming "token humans". Our recommendations for making the human oversight process more useful is to only appoint people who understand the system well and are able to communicate any potential risks effectively with a variety of stakeholders so that they can take the appropriate action.

9. **In addition to the existing EU legislation, in particular the data protection framework, including the General Data Protection Regulation and the Law Enforcement Directive, or, where relevant, the new possibly mandatory requirements foreseen above (see question above), do you think that the use of remote biometric identification systems (e.g. face recognition) and other technologies which may be used in public spaces need to be subject to further EU-level guidelines or regulation:**

No further guidelines or regulations are needed Biometric identification systems should be allowed in publicly accessible spaces only in certain cases or if certain conditions are fulfilled (please specify)
Other special requirements in addition to those mentioned in the question above should be imposed (please specify)
**[x] Use of Biometric identification systems in publicly accessible spaces, by way of exception to the current general prohibition, should not take place until a specific guideline or legislation at EU level is in place**.
Biometric identification systems should never be allowed in publicly accessible spaces
No opinion

We had the chance to discuss this question at length when the Montreal AI Ethics Institute held our [public meetups](#) on the European Commission's AI Whitepaper. We began by addressing the new challenges regarding contact-tracing and covid-19. A distinction was then made between using the data collected by contact-tracing apps to predict or anticipate where people will go and potentially catch or transmit the virus, or whether the app was used to steer individuals away from red zones. One participant then suggested that the more important question was about who gets to define the parameters around the exceptional collection of data in these unique times. For example, how is the data stored or processed, who has access to it? No legislation had really anticipated the situation the pandemic would put us in with regards to data privacy and security. Surrounding this matter, concerns were also expressed about future uses of data. When the pandemic is over, for example, there seemed to be very little faith that governments and big tech companies were going to simply delete the data. In addition, participants were not convinced by the narrative that pits privacy against the public good in the context of the pandemic and of contact-tracing apps. It was suggested that the scientific validity of the tools used was much more germane to the debate. While this discussion is not specifically about biometric identification systems, we believe the core ideas expressed are nonetheless relevant to these kinds of systems. Thus, we hold that *if* biometric identification systems are used, they must be able to fulfil the purpose for which they are implemented while also, by far, being the best way of going about the task. The risks present in the use of biometric identification systems, paired with participants' perception that governments and companies cannot be trusted with citizens' data, led us to be very skeptical of the use of biometric identification systems.

Ultimately, one participant suggested that we must frame the question around the regulation of facial recognition as follows: "What is a good reason *not* to regulate facial recognition?" This places the burden on the other side, so to speak, and asks those who believe it does *not* require regulation to make a compelling argument. We agree with this participant, and hence believe that no biometric identification systems or similar technologies should be used in public spaces until further EU regulation is put in place.

*Please specify your answer*

*Please see the attached PDF for more information.*

10. **Do you believe that a voluntary labelling system (Section 5.G of the White Paper) would be useful for AI systems that are not considered high-risk in addition to existing legislation?**

Very much
Much **[x] Yes, but must be backed by economic incentives. (i.e. SECure certification as proposed by MAIEI)**
Rather not
Not at all
No opinion

We believe that a voluntary labelling system could be useful for systems that are not considered high-risk. However, we would privilege compulsory measures even for low or medium risk AI systems (as was detailed in our answer to the previous question). If a voluntary labelling system were to be implemented, we believe it should be supported by strong economic incentives. This is what we at the Montreal AI Ethics Institute have attempted in proposing the SECure framework, which relies in great part on consumers choosing AI systems which qualify for the SECure certificate.[40] We hope that having AI systems that are certified as environmentally and socially responsible will prompt consumers to seek these certified systems, which will then incite AI companies to build more socially and environmentally responsible AI systems. Hence we believe a similar voluntary labeling mechanism that creates an economic incentive for companies can be helpful in regulating AI.

*Do you have any further suggestions on a voluntary labelling system? 500 character(s) maximum*

*Please see the attached PDF for more information.*

### 11. What is the best way to ensure that AI is trustworthy, secure and in respect of European values and rules?

Compliance of high-risk applications with the identified requirements should be self-assessed ex-ante (prior to putting the system on the market)
Compliance of high-risk applications should be assessed ex-ante by means of an external conformity assessment procedure
Ex-post market surveillance after the AI-enabled high-risk product or service has been put on the market and, where needed, enforcement by relevant competent authorities
**A combination of ex-ante compliance and ex-post enforcement mechanisms [x]**
Other enforcement system
No opinion

The biggest hurdle as we identified in a previous answer was that we need to have practical implements in place that help to translate principles into action. Specifically, we advocate for having verifiable claims that can demonstrate both the trustworthiness of the inputs and the outputs along with the models and other components of the AI development and deployment lifecycle. Specifically, just as it is the case with cybersecurity, maintain the trustworthiness and security of an AI system requires an iterative and constant evaluation approach that follows up with the designers, developers, and management to know that the system continues to maintain its state of being trustworthiness after being modified through its interactions with real-world data.

As for respecting the European values and rules, we advocate that the best results lie with the locally empowered citizens who are closest to the problem and hence have the most cultural and contextual knowledge that they can apply to build solutions that meaningfully address their situation while giving the community dignity and the ability to work with each other to create solutions for themselves.

---

[40] Gupta, A., Lanteigne, C., & Kingsley, S. (2020). SECure: A Social and Environmental Certificate for AI Systems. arXiv preprint arXiv:2006.06217.

*Do you have any further suggestions on the assessment of compliance? 500 character(s) maximum*

*Please see the attached PDF for more information.*

**Section 3: Safety and liability implications of AI, IoT and robotics**

**12. The current product safety legislation already supports an extended concept of safety protecting against all kinds of risks arising from the product according to its use. However, which particular risks stemming from the use of artificial intelligence do you think should be further spelled out to provide more legal certainty?**

Cyber risks
Personal security risks
**Risks related to the loss of connectivity [x]**
**Mental health risks [x]**

We consider that risks related to the loss of connectivity and mental health risks would greatly benefit from being better spelled out with regards to legal certainty. This is because these risks are particularly new in comparison to the two other types of risks mentioned above. Their novelty means they are not (or at least not adequately) recognized by the law, and hence these risks are particularly likely to materialize into significant harm.

As covered elsewhere in our answers, cyber risks and personal security risks should really be captured and integrated deeply into the AI development life cycle. Legal certainty can make a potential difference here but we believe empowered and well-informed citizens can become stronger agents of change who push for the adoption of responsible AI principles.

*In your opinion, are there any further risks to be expanded on to provide more legal certainty? 500 character(s) maximum*

*Please the attached PDF for more information.*

**13. Do you think that the safety legislative framework should consider new risk assessment procedures for products subject to important changes during their lifetime?**

**[x] Yes**
No

Online learning settings where the machine learning system changes in response to its interactions with the real-world data. In the context of machine learning security, as detailed before, there are a wide variety of attacks that can be mounted on the system that can compromise its confidentiality, integrity, and availability requires new risk assessment procedures that can concretely identify the places where the systems can fail and assign scores and severity levels, following the model of the CVSS[41] so that there is some standardization in this work across the industry and comparisons can be made in doing risk assessments.

Additionally, having such a new mechanism be structured as a complement to existing risk scoring and assessment mechanisms in the domain of cybersecurity will be important. Particularly when it comes to practical integration into the larger software development and deployment lifecycle so that friction in use is minimized. Data and Privacy Impact Assessments provide a good base to start and also have the advantage of being something familiar that developers and information security have an understanding of so they are more likely to adopt a new solution that either exists as a separate item that they fill or it becomes an extension of the existing assessment tools.

*Do you have any further considerations regarding risk assessment procedures? 500 character(s) maximum*

**14. Do you think that the current EU legislative framework for liability (Product Liability Directive) should be amended to better cover the risks engendered by certain AI applications?**

**Yes [x]**
No
No opinion

As mentioned in the previous response, there are emergent risks from the integration of AI components into the software infrastructure pipeline which opens up new risks that need to be categorized and better managed such that liability is more defined and can be applied confidently in cases where there is a clear need without having to invent ideas on-the-fly in terms of how novel situations will have to be addressed. Sector- and application-specific guidance would be required because of the different nature of the risks that emerge from these AI systems, for example the kind of attacks and failures that happen in a natural language processing system are different from those in a computer vision system and hence require nuance be baked into the legislative framework in the form of flexibility of interpretation and assessment such that liability is well covered in the different cases and applications of AI.

---

[41] *Common Vulnerability Scoring System SIG*. (n.d.). FIRST — Forum of Incident Response and Security Teams. Retrieved 14 June 2020, from https://www.first.org/cvss

*Do you have any further considerations regarding the question above? 500 character(s) maximum*

*Please see the attached PDF for more information.*

### 15. Do you think that the current national liability rules should be adapted for the operation of AI to better ensure proper compensation for damage and a fair allocation of liability?

**Yes, for all AI applications [x]**
Yes, for specific AI applications
No
No opinion

We believe that adapting the current national liability rules for all AI applications would be beneficial in helping ensure that liability is fairly allocated. This is because, for one, the introduction of AI into products like cars or loan management systems creates new challenges in terms of liability. While one may hold the bank liable if its employees discriminate on the basis of gender or race in granting loans, liability becomes much more difficult to assign if an AI system is the one making these decisions. Should the developers of the AI system be held liable? Or should the bank, since it decided to purchase the system? Should both parties (or none of them) be held liable in this case? Perhaps another actor may be liable for the biased decisions made in affording loans to applicants. A similar dilemma with regard to liability arises in the case where a person is harmed or killed by an autonomous vehicle. In light of difficult situations like this, we believe national liability rules should be adapted for all AI applications.

In relation to these issues, we also wonder how to properly reason about the unintentional harms that might arise from an AI system. It may be fruitful to investigate a way through which organizations can demonstrate that they have taken all the necessary measures to prevent their AI system from causing harm, and are hence protected from being held liable for any damages that might arise from unintended behaviour as the system interacts with real-world data. While this can be helpful in determining liability, it is important that such a framework be extremely robust, and not serve as a quick and easy way to rid oneself of their responsibilities when building or using an AI system.

*Do you have any further considerations regarding the question above? 500 character(s) maximum*

*Please see the attached PDF for more information.*